\documentclass{article}
\usepackage[margin=1in]{geometry}
\usepackage{graphicx} 
\usepackage{endfloat} 
\usepackage{amsfonts, amsmath} 
\usepackage{accents} 
\usepackage{booktabs}
\usepackage{hyperref}

\usepackage{url}
\title{Rank-based estimators of global treatment effects for cluster randomized trials with multiple endpoints}
\author{Davies Smith, Emma$^{1,2, *}$ \and Jairath, Vipul$^{2,3,4}$ \and Zou, Guangyong$^{2,3,5}$}
\date{\footnotesize
    $^1$Biostatistics, Harvard T.H. Chan School of Public Health, Boston, MA, USA\\%
    $^2$Epidemiology and Biostatistics, Schulich School of Medicine \& Dentistry, Western University \\
    $^3$Alimentiv, Inc., London, ON\\
	$^4$Gastroenterology, Department of Medicine, Schulich School of Medicine \& Dentistry, Western University \\
	$^5$Robarts Research Institute, Schulich School of Medicine \& Dentistry, Western University \\
	$^{*}$Corresponding author: Emma Davies Smith, esmith@hsph.harvard.edu
    \\[2ex]%
    \normalsize
    \today
}

\begin{document}
\maketitle
\section*{Abstract}
Cluster randomization trials commonly employ multiple endpoints. When a single summary of treatment effects across endpoints is of primary interest, global hypothesis testing/effect estimation methods represent a common analysis strategy. However, specification of the joint distribution required by these methods is non-trivial, particularly when endpoint properties differ. We develop rank-based interval estimators for a global treatment effect referred to as the ``global win probability,” or the probability that a treatment individual responds better than a control individual on average. Using endpoint-specific ranks among the combined sample and within each arm, each individual-level observation is converted to a ``win fraction" which quantifies the proportion of wins experienced over every observation in the comparison arm. An individual’s multiple observations are then replaced by a single ``global win fraction," constructed by averaging win fractions across endpoints. A linear mixed model is applied directly to the global win fractions to recover point, variance, and interval estimates of the global win probability adjusted for clustering. Simulation demonstrates our approach performs well concerning coverage and type I error, and methods are easily implemented using standard software. A case study using publicly available data is provided with corresponding R and SAS code.

\section*{Keywords}
Cluster randomized trials; Global treatment effects; Nonparametric rank-sum test; Win ratio; U-statistics; Linear mixed models
\pagebreak

\section{Introduction}
Complex diseases with complex interventions demand complex trials. Cluster randomized trials, or cluster trials, can simplify the delivery of complex interventions and mitigate the risk of contamination by randomly allocating groups of individuals, or ``clusters," rather than individuals to intervention arms. However, their design and analysis is complicated by correlation among individuals within the same cluster, or ``intracluster correlation." \cite{crts} Correlation structures are complicated even further when multiple endpoints are of primary interest. In addition to multiple sources of intracluster correlation which may differ by endpoint, endpoint correlations within-subject must also be captured.

Multiple endpoints are commonly associated with multiplicity, but they may be employed for a variety of reasons, resulting in different implications for analysis.\cite{fda_endpts} In this paper, we focus on the scenario where a single summary of the treatment effect across multiple endpoints, or a ``global treatment effect,” is of primary interest. This scenario may arise when there is a lack of consensus on which single best endpoint to use, or when multiple endpoints are necessary to characterize disease burden at a single time point, risk-benefit trade-off, or even longitudinal disease course.\cite{fda_endpts,  ob_interpret} Global tests represent a common analytic approach in this scenario, and assess overarching hypotheses about all endpoints simultaneously using a single test statistic constructed from their joint distribution. By testing endpoints jointly rather than separately, multiplicity is not a concern, and power is often greater than the multiple corresponding univariate tests.\cite{ristl} 



With $K$ endpoints, $K$-df global tests assess $K$ hypotheses. Hotelling’s $T^2$ test belongs to the family of $K$-df global testing methods, and assesses the two-sided alternative of a non-zero mean difference for at least one endpoint. Since a two-sided alternative does not consider effect direction, the null hypothesis of no effect for all endpoints may be rejected when treatment is harmful for any or all endpoints. When designing a trial, there is often some \textit{a priori} knowledge about effect direction, particularly when endpoints represent alternative measures of the same phenomena or repeated measurements on the same individual. Treatment benefit for all endpoints may also be of primary interest, \textit{i.e.}, in a superiority trial. In these scenarios, Hotelling's $T^2$ test is inappropriate and can lead to diminished power.\cite{obrien} Specification of the correlation structure required by $K$-df tests is also non-trivial, particularly when endpoint types or scales differ.\cite{yoon}

When the global treatment effect is of primary interest, the $K$ hypotheses are reduced to a single hypothesis regarding the global treatment effect alone, and a $1$-df global test results. Motivated by a need to obtain a single probability statement regarding multiple disparate endpoints among few subjects, O'Brien\cite{obrien} developed three 1-df global tests of a directed alternative: the ordinary least squares (OLS) test, the generalized least squares (GLS) test, and the nonparametric rank-sum test. All three tests construct the global treatment effect as the sum (or mean) of the endpoint-specific effects, but differ in their standardization and weighting schemes. Each of the tests is actually equivalent to: (1) a univariate test of the global treatment effect; (2) a combination of the univariate, endpoint-specific test statistics;\cite{lachin2014} and notably, (3) a two-sample $t$-test of a composite endpoint constructed as the within-subject mean of standardized responses.\cite{tamhane} Thus, if an appropriate standardization can be identified and the corresponding composite is relevant, univariate tests may be applied directly, avoiding explicit specification of complex correlation structures.

The ``parametric" OLS and GLS tests apply $Z$-standardization to each endpoint under the assumption that they are multivariate normal, but this assumption is often untenable. O'Brien's ``nonparametric" rank-sum test instead replaces responses with their rank in the pooled sample.\cite{obrien} Ranks are then summed across endpoints within-subject, yielding composite ``rank-sums." For large sample sizes, correlation between rank-sums is sufficiently weak to permit application of the Central Limit Theorem. Thus, the nonparametric rank-sum test is constructed as a $t$-test for the mean difference in rank-sums, here the global treatment effect. Of the three tests, O'Brien promoted the rank-sum test for general use, as it suffers from little loss in efficiency when parametric assumptions are met and provides gains in power otherwise.\cite{obrien, sankoh} While developed with one-sided alternatives in mind, the nonparametric rank-sum test is also applicable to two-sided alternatives.\cite{huang2005, huang2008}

The nonparametric rank-sum test suffers from three main drawbacks. First, the reporting of hypothesis tests and p-values, while informative in their own right, is no longer sufficient. Trials are now strongly encouraged to report point estimates and their uncertainty, preferably in the form of confidence intervals.\cite{consort}  Second, the mean difference in rank-sums is not an easily interpretable measure of group differences. However, as we demonstrate in what follows, it may be re-expressed in terms of ``the win probability." The win probability is a nonparametric treatment effect defined as the probability that a randomly selected treatment individual responds better than, or ``wins over," a randomly selected control individual. The win probability and related win-based measures are gaining traction as clinically relevant effect measures with desirable statistical properties.\cite{acion} Third, there are currently no feasible extensions to the cluster randomized setting.

Zou\cite{zou} recently developed rank-based interval estimators for a single win probability within cluster randomized trials by transforming individual-level responses into ``win fractions." Win fractions are proportions that summarize the wins experienced by an individual when compared to all others within the comparator arm. For example, a win fraction of 0.75 suggests that a (treatment) individual responded better than, or ``won over," 75\% of those in the comparator (control) arm. Zou\cite{zou} demonstrated that unbiased and consistent estimators of the win probability could be recovered from a linear mixed model for the win fractions with a random cluster intercept. Simulation studies demonstrated that resulting confidence intervals perform well, maintaining nominal coverage probabilities and type I error rates. Perhaps most importantly, these methods can be easily implemented using standard software and place win-based treatment effects within a regression framework.

This paper presents an easily implemented, widely applicable, and interpretable solution to the otherwise complex analysis of cluster randomized trials with multiple endpoints by building upon the ideas of O'Brien's nonparametric rank-sum test\cite{obrien} and the mixed model estimators of Zou\cite{zou}. Using ranks, a composite ``global win fraction" is first constructed for each individual within the cluster trial as the within-subject mean of their endpoint-specific win fractions. A univariate linear mixed model is then applied directly to the global win fractions to obtain point and variance estimates for the global win probability adjusted for intracluster correlation. While hypothesis testing methods are provided, emphasis is placed on confidence interval estimation. The developed methods are simple yet flexible enough to handle multiple endpoints with differing properties such as type, scale, or priority, avoid explicit specification of complex correlation structures, can be implemented using existing software, yield meaningful treatment effects, and have the potential to increase power -- a particular concern of cluster trials.

The rest of the paper is organized as follows. Section 2 defines notation, reviews mixed model estimators for a single win probability, provides a straightforward extension to the global win probability for multiple endpoints, and demonstrates translation to alternative win measures and equivalence with the mean difference in rank-sums. In Section 3, simulation studies based on design parameters encountered in practice demonstrate the ability of the developed methods to maintain interval coverage and the type I error rate while providing high power, even when relatively few clusters are randomized. In Section 4, a case study using publicly available data from the SHARE\cite{share} cluster trial exemplifies the application of the global win probability to two endpoints with different types and priority. Corresponding SAS and R code is provided in the Appendix to assist with implementation. The paper closes with a discussion, including a sketch of how sample size estimation may be conducted at present. 

\section{Methods}
\subsection{Notation, context, and assumptions}
Consider a two-arm cluster randomized trial allocating $C_0$ and $C_1$ clusters to the control $(i=0)$ and treatment $(i=1)$ arms, respectively. Let $c=1, 2, \hdots, C_i$ index the clusters randomized to the $i^{th}$ intervention arm and $j=1, 2, \hdots, n_{ic}$ index the $n_{ic}$ individuals within the $ic^{th}$ cluster. All $n_{ic}$ individuals receive the intervention allocated to their cluster. Then, $C=C_0+C_1$ is the total number of clusters randomized by the trial, $N_{i} = \sum_{c=1}^{C_i} n_{ic}$ is the total number of individuals within the $i^{th}$ intervention arm, and $N=N_0+N_1$ is the total number of individuals within the trial. Due to randomization, clusters are independent within and between arms. 

Suppose $k=1,2,\hdots, K$ endpoints are recorded for each individual within the trial. Let $X_{icjk}$ denote the $k^{th}$ response of the $j^{th}$ individual within the $ic^{th}$ cluster. Assume all individuals are completely observed so that $(N \times K)$ total responses are recorded. The $X_{icjk}$ are not independent within the ${ic}^{th}$ cluster, but are assumed to be identically distributed according to non-degenerate distribution function $F_{ik}$. To accommodate discrete endpoints, $F_{ik}$ is defined as the ``normalized distribution function," $F_{ik}(x) = 0.5 \left[ F_{ik}(x)^{-} + F_{ik}^{+}(x) \right]$
where $F_{ik}^{-}(x) = \Pr(X_{icjk} < x)$ and $F_{ik}^{+}(x) = \Pr(X_{icjk} \leq x)$ are the left- and right-continuous distribution functions, respectively. Finally, assume endpoints are at least ordinal in nature so any two responses can be ordered, and without loss of generalizability, greater responses correspond to better health.

\subsection{Global win probability, and related measures}
%


Following Zou,\cite{zou} we consider the individual-level win probability for a single endpoint, hereon referred to simply as the win probability, which compares individual-level responses between arms rather than cluster-level summaries. For the $k^{th}$ endpoint, the win probability takes the form,
\begin{equation}
\theta_{k} = \Pr(\underbrace{X_{1cjk} > X_{0cjk}}_{\text{``Win"}})+ 0.5 \Pr(\underbrace{X_{1cjk} = X_{0cjk}}_{\text{``Tie"}}).
\label{eq: theta_k}
\end{equation}
The win probability is equivalent to the Mann-Whitney U test statistic and has had over a dozen unique names including the area under the receiver operating characteristic curve, concordance or the c-index, the probabilistic index, and the common language effect size.\cite{zou_stroke} It is also a unifying link between commonly encountered effect measures. Of particular note, when the $k^{th}$ endpoint is normally distributed, $\theta_k$ is a one-to-one function of the standardized mean difference or Cohen's effect size $\delta$, with $\theta_{k} = \Phi(\delta/\sqrt{2})$ where $\Phi(\cdot)$ is the standard normal distribution function. Cohen's qualitative benchmarks are then easily transferable to the win probability, with $\theta_{k} = 0.5, 0.56, 0.64,$ and $0.71$ corresponding to null, small, moderate, and large effects, respectively.\cite{effectsizes}

When $K$ endpoints are of joint interest, we propose that the average of the $K$ corresponding win probabilities serves as the global treatment effect. Formally, the ``global win probability" is defined as,
\begin{equation}
\theta = \frac{\sum_{k=1}^{K} w_{k} \theta_{k}}{\sum_{k=1}^{K} w_{k}}
\label{eq: wt_theta}
\end{equation}
where $w_{k}$ represents the contribution weight of the $k^{th}$ endpoint. In what follows, we focus primarily on equal weights for simplicity, $w_{k} = 1/K$ for all $k$, in which case the global win probability reduces to the simple average,
\begin{equation*}
\theta = \frac{1}{K} \sum_{k=1}^{K} \theta_{k}.
\label{eq: theta}
\end{equation*}
The global win probability may be formally interpreted as the probability that a randomly selected individual from a treatment cluster responds no worse than a randomly selected individual from a control cluster with respect to the $K$ endpoints, on average. Alternatively, multiplying $\theta$ by the number of endpoints $K$ provides the expected number of endpoints on which a randomly selected individual from a treatment cluster will respond no worse than a randomly selected individual from a control cluster.\cite{ob_interpret}

\subsubsection{Global win difference}
Rather than the global win probability, one may consider the global win difference. For a single endpoint, the win difference has also been referred to as Somer's $d$, the Mann-Whitney difference,\cite{lachin} and the proportion in favor of treatment.\cite{buyse} The win difference is commonly used for methodology development as it avoids explicit consideration of tie probabilities, $\Pr(X_{1cjk} = X_{0cjk})$. For example, the global win difference was employed by Huang et al when developing improvements of the nonparametric rank-sum test,\cite{huang2005, huang2008} and by Lachin when developing multivariate distribution-free hypothesis testing methods.\cite{lachin}

First, note that the $k^{th}$ win probability presented within Equation (\ref{eq: theta_k}) is defined with respect to a treatment win, $(X_{1cjk} > X_{0cjk})$. The probabilities of a treatment win, loss, and tie sum to unity,
\begin{equation}
\Pr(X_{1cjk} > X_{0cjk}) + \Pr(X_{1cjk} < X_{0cjk}) + \Pr(X_{1cjk} = X_{0cjk}) = 1.
\label{eq: unity}
\end{equation}
Since the win probability distributes ties equally between arms and a loss for treatment is a win for control, $(X_{0cjk} > X_{1cjk})$, it follows from Equation (\ref{eq: unity}) that the $k^{th}$ win probability defined with respect to a control win, or the ``control win probability," is the complement of the treatment win probability, $ 1 - \theta_{k}$. 

The $k^{th}$ win difference is defined as the difference between the treatment and control win probabilities,
\begin{equation}
\Delta_{k} = \Pr(X_{1cjk} > X_{0cjk}) - \Pr(X_{1cjk} < X_{0cjk})
\label{eq: Delta_k}
\end{equation}
or equivalently, $\Delta_{k} = \theta_{k} - (1-\theta_k)$. $\Delta_{k}$ represents a generalization of the risk difference for binary endpoints to all endpoint types, and ranges between $-1$ and $+1$ with $\Delta_k = 0$ suggesting no treatment effect and $\Delta_k > 0$ treatment benefit relative to the control. With $K$ endpoints, the ``global win difference" may be obtained in a similar fashion to Equation (\ref{eq: wt_theta}) as $\Delta = \sum_{k} w_k \Delta_k / \sum_{k} w_k$. Thus, estimators provided in Sections \ref{ss: pv_est} and \ref{ss: i_est} may be translated to the global win difference by applying the transformations $\hat{\Delta} = 2 \hat{\theta} -1$ and $\widehat{\text{Var}}(\hat{\Delta}) = 4 \widehat{\text{Var}}(\hat{\theta})$.

\subsubsection{Global win odds}
Agresti proposed the generalized odds ratio, commonly referred to as ``Agresti's $\alpha$," as a treatment effect for ordinal endpoints.\cite{agresti} For the $k^{th}$ endpoint, $\alpha_k$ is equal to the ratio of the probability of a treatment win to a control win,
\begin{equation*}
\alpha_k = \frac{\Pr(X_{1cjk} > X_{0cjk})}{\Pr(X_{1cjk} < X_{0cjk})}
\end{equation*}
and is equivalent to the odds ratio when the $k^{th}$ endpoint is binary. Pocock et al popularized the use of Agresti's $\alpha$ as a measure of effect size for prioritized time-to-event composites within cardiovascular trials,\cite{winratio} referring to it as the ``win ratio." For a single continuous survival endpoint, $\alpha_k$ is equivalent to the inverse of the hazard ratio when the proportional hazards assumption holds. However, $\alpha_k$ excludes the probability of ties which can be both informative and substantial for discrete endpoints. 

Dong et al\cite{dong} and Brunner et al\cite{winodds} instead advocate for the use of the ``win odds" which incorporates ties and is equal to the ratio of the treatment and control win probabilities. For the $k^{th}$ endpoint,
\begin{equation*}
\lambda_k = \frac{\Pr(X_{1cjk} > X_{0cjk}) + 0.5 \Pr(X_{1cjk} = X_{0cjk})}{\Pr(X_{1cjk} < X_{0cjk}) + 0.5 \Pr(X_{1cjk} = X_{0cjk})}
\end{equation*}
or equivalently, $\lambda_k = \theta_k / (1-\theta_k)$.When the $k^{th}$ endpoint is continuous, $\Pr(X_{1cjk} = X_{0cjk}) = 0$ so that $\lambda_k = \alpha_k$. The win odds $\lambda$ range between $0$ and $+\infty$ with $\lambda = 1$ indicating no treatment effect and $\lambda > 1$ treatment benefit relative to the control. With $K$ endpoints, the ``global win odds" are defined as $\lambda = \sum_{k} w_k \lambda_k / \sum_{k} w_k$. Thus, estimators provided in Sections \ref{ss: pv_est} and \ref{ss: i_est} may be translated to the global win odds through application of the $\delta$-method, with $\hat{\lambda} = \hat{\theta} / (1-\hat{\theta})$ and $\widehat{\text{Var}}(\ln \hat{\lambda}) = \widehat{\text{Var}}(\hat{\theta})/[\hat{\theta}(1-\hat{\theta})]^2$.

\subsection{Global win fractions} \label{ss: gwf}
Point and variance estimation for associated measures such as the Wilcoxon Mann-Whitney U-test statistic or win difference have traditionally relied on the construction and comparison of all $N_1 N_0$ pairs consisting of one treatment and control response.\cite{buyse, mw_test} However, the construction of all ``pairwise comparisons" is computationally expensive, even for moderate sample sizes. We focus instead on novel estimators which transform each individual's observed response into a proportion summarizing their wins and ties experienced, referred to as a ``win fraction."\cite{zou, zou_stroke, zou_pharma, zou_contemp} Unlike all pairwise comparisons, the number of win fractions is equal to the number of individuals within the cluster trial.


Formally, the $k^{th}$ win fraction for the $j^{th}$ individual in the $c^{th}$ treatment cluster is provided by,
\begin{equation}
Y_{1cjk} = \frac{1}{N_0} \sum_{c'=1}^{C_0} \sum_{j'=1}^{n_{0c'}} H(X_{1cjk} - X_{0c'j'k})
\label{eq: y_1cjk}
\end{equation} 
where $H(X_{1cjk} - X_{0c'j'k})$ is the Heaviside function taking the value +1 when $(X_{1cjk} > X_{0c'j'k})$ or a ``win" occurs, +0.5 when $(X_{1cjk} = X_{0c'j'k})$ or a ``tie" occurs, and +0 when $(X_{1cjk} < X_{0c'j'k})$ or a ``loss" occurs. In other words, win fractions are equal to the within-subject mean of the $(N-N_i)$ pairwise comparisons involving $X_{icjk}$. Of particular note, $Y_{1cjk}$ may also be expressed as $\hat{F}_{0k}(X_{1cjk})$, where $\hat{F}_{ik}(x)$ is the empirical distribution function (ECDF) of the $k^{th}$ endpoint in the $i^{th}$ arm. That is, $Y_{1cjk}$ is the percentile that treatment observation $X_{1cjk}$ occupies among all control observations.\cite{zou} Similarly, for the $j^{th}$ individual in the $c^{th}$ control cluster,
\begin{equation}
Y_{0cjk} = \frac{1}{N_1} \sum_{c'=1}^{C_1} \sum_{j=1}^{n_{1c'}} H(X_{0cjk} - X_{1c'j'k}) 
\label{eq: y_0cjk}
\end{equation}
or $Y_{0cjk} = \hat{F}_{1}(X_{0cjk})$, the percentile that $X_{0cjk}$ occupies among all treatment observations. 


As a result of this relationship with the ECDF, win fractions may be conveniently expressed in terms of ranks. With ties, ``ranks" refers to ``midranks" or the average of the tied positions. Following Hoeffding,\cite{hoeffding} define two ranks for individual-level treatment response $X_{1cjk}$: (1) the ``overall rank" among all $N$ responses in the trial,
$$ R_{1cjk} = 0.5 + \sum_{c'=1}^{C_{1}} \sum_{j'=1}^{n_{1c'}} H(X_{1cjk} - X_{1c'j'k}) + \sum_{c'=1}^{C_{0}} \sum_{j'=1}^{n_{0c'}} H(X_{1cjk} - X_{0c'j'k})$$
or equivalently, $R_{1cjk} = 0.5 + N_1 \hat{F}_1(X_{1cjk}) + N_0 \hat{F}_{0}(X_{1cjk})$, and (2) the ``group-specific rank" among the $N_{1}$ responses in the treatment arm,
$$ G_{1cjk} = 0.5 + \sum_{c'=1}^{C_1} \sum_{j'=1}^{n_{1c}} H(X_{1cjk} - X_{1c'j'k})$$
or equivalently, $G_{1cjk} = 0.5 + N_1 \hat{F}_{1}(X_{1cjk})$. These ranks can be defined analogously for control observations. It then follows from Equations (\ref{eq: y_1cjk}) and (\ref{eq: y_0cjk}) that the win fractions may be expressed generally as,
\begin{equation}
Y_{icjk} = \frac{R_{icjk} - G_{icjk}}{N-N_i}.
\label{eq: rank_y}
\end{equation}
The rank-based form of the win fractions simplifies and speeds up calculation significantly as all $N_{1} N_{0}$ pairwise comparisons do not need to be constructed, rather only three sets of ranks. 

To estimate the global win probability, a single ``global win fraction" is constructed for each individual as the (weighted) mean of their $K$ endpoint-specific win fractions,
\begin{equation}
\bar{Y}_{icj\cdot} = \frac{\sum_{k=1}^{K} w_{k} Y_{icjk}}{\sum_{k=1}^{K} w_k}. 
\label{eq: gwf}
\end{equation}
With equal endpoint weights, Equation (\ref{eq: gwf}) reduces to $\bar{Y}_{icj\cdot} = \sum_{k=1}^{K} Y_{icjk}/K$, or the simple within-subject mean of the $K$ win fractions. Transformation of the multiple responses into global win fractions ensures that endpoints share a common support and contribute proportionally to the treatment effect. Global win fractions are also interpretable at the individual level as the average proportion of responses exceeded (or tied) in the comparator arm, unlike the mean of normal deviates or the rank-sum, and permit any combination of binary, ordinal, count, or continuous endpoints.

\subsection{Mixed model estimators} \label{ss: pv_est}
Several point and variance estimators have been developed for independent observations using quantities similar or equal to win fractions. Sen\cite{sen1960, sen1967} developed variance estimators for the win probability by decomposing its U-statistic estimator into identically distributed and asymptotically uncorrelated ``structural components," one-to-one functions of win fractions. Arvensen\cite{arvensen} derived similar variance estimators for U-statistics by applying the leave-one-out (LOO) jackknife, and Hanley and Hajian-Tilaki\cite{hanleyhajian} demonstrated the equivalence of LOO jackknife pseudo-observations and win fractions. DeLong et al\cite{delong} extended Sen's structural component method\cite{sen1960} to estimate the covariance of multiple AUCs, or multiple win probabilities.  Brunner and Munzel\cite{brunmun} provided a rank-based solution to the nonparametric Behrens-Fisher problem by framing hypotheses with respect to the win probability and constructing estimators using win fractions.

Zou\cite{zou} extended these ideas and estimators to cluster randomized trials, demonstrating that an unbiased estimator of a single win probability $\theta_{k}$ can be obtained as the mean treatment win fraction, $\hat{\theta}_{k} = \bar{Y}_{1\cdot\cdot k}$, or since a win for control is a loss for treatment, one minus the mean control win fraction, $\hat{\theta}_{k} = 1-\bar{Y}_{0\cdot\cdot k}$. It was also shown that asymptotically, $\hat{\theta}_{k} \sim \mathcal{N}(\theta_k, \text{Var}(\hat{\theta}_k))$ where $\text{Var}(\hat{\theta}_{k}) \approx \text{Var}(\bar{Y}_{1\cdot\cdot k}) + \text{Var}(\bar{Y}_{0\cdot \cdot k})$. Two estimators of the $\bar{Y}_{i\cdot \cdot k}$ and a total of three corresponding variance estimators, $\widehat{\text{Var}}(\bar{Y}_{i\cdot\cdot k})$, were investigated. Here, we construct a single transformed response for each individual within the cluster randomized trial, \textit{i.e.}, a global win fraction. In what follows, we obtain point and variance estimators of the global win probability by applying these univariate methods, the mixed model estimators specifically, directly to the global win fractions.

\subsubsection{Application to global win fractions}
An unbiased estimator of the global win probability $\theta$ and a consistent estimator of its variance can be obtained by applying the following linear mixed model to the global win fractions,
\begin{equation}
\bar{Y}_{icj\cdot} = \beta_0 + \beta_1 \text{Arm}_i + \alpha_{ic} + \varepsilon_{icj}
\label{mod}
\end{equation}
where $\text{Arm}_{i} = \mathbb{I}(i=1)$ is an indicator equal to 1 if individual $icj$ belongs to a treatment cluster and $0$ if control, $\alpha_{ic} \stackrel{i.i.d.}{\sim} \mathcal{N}(0, \sigma^2_{\alpha})$ represents the random intercept of the $ic^{th}$ cluster, $\varepsilon_{icj} \stackrel{i.i.d.}{\sim} \mathcal{N}(0, \sigma^2_{\varepsilon})$ represents the residual of the $icj^{th}$ win fraction, and $\alpha_{ic}$ and $\varepsilon_{icj}$ are assumed to be independent. From these definitions it follows that $\hat{\beta}_1 = (\bar{Y}_{1\cdot \cdot \cdot} - \bar{Y}_{0 \cdot \cdot \cdot}) = 2 \hat{\theta}-1$ and an estimator of the global win probability can be recovered from the fitted model as $\hat{\theta} = 0.5 (\hat{\beta}_1 + 1)$. Since the two intervention arms are independent, $\widehat{\text{Var}}(\hat{\beta}_1) = \widehat{\text{Var}}(\bar{Y}_{1\cdot \cdot \cdot}) + \widehat{\text{Var}}(\bar{Y}_{0\cdot \cdot \cdot})$ and thus $\widehat{\text{Var}}(\hat{\theta}) = \widehat{\text{Var}}(\hat{\beta}_1)$. An estimate of the intracluster correlation of the global win fractions is also be obtained as $\hat{\rho} = \hat{\sigma}^2_{\alpha}/(\hat{\sigma}^2_{\alpha} + \hat{\sigma}^2_{\varepsilon})$.

When the variance components $\sigma_{\alpha}^2$ and $\sigma_{\varepsilon}^2$ are known, $\hat{\beta}_0$ and $\hat{\beta}_1$ are equivalent to both weighted least squares (WLS) and generalized least squares (GLS) estimators. Thus, the resulting mixed model estimator of the $i^{th}$ mean win fraction is, 
\begin{equation}
\bar{Y}_{i\cdot \cdot \cdot} = \frac{\sum_{c=1}^{C_i} w_{ic} \bar{Y}_{ic\cdot \cdot}}{\sum_{c=1}^{C_i} w_{ic}}
\label{eq: gls_mean}
\end{equation}
where $\bar{Y}_{ic\cdot \cdot} = \sum_{j=1}^{n_{ic}} \bar{Y}_{icj\cdot} / n_{ic}$ is the sample mean of the global win fractions within the $ic^{th}$ cluster. The weights for each cluster are given by,
\begin{equation*}
w_{ic} = \left( \frac{\sigma^2}{n_{ic}} [ 1 + (n_{ic} - 1) \rho ] \right)^{-1}
\end{equation*}
where $\sigma^2 = (\sigma^2_{\alpha} + \sigma^2_{\varepsilon})$, or the ``total variation" of the global win fractions, and $\rho = \sigma^2_{\alpha} / \sigma^2$ represents their intracluster correlation. Finally, $\text{Var}(\bar{Y}_{i\cdot \cdot \cdot}) = 1/(\sum_{c=1}^{C_i} w_{ic} )$. In reality, the variance components are unknown, resulting in more complex feasible generalized least squares (FGLS) estimators when replaced by $\hat{\sigma}_{\alpha}^2$ and $\hat{\sigma}_{\varepsilon}^2$. However, both GLS and FGLS estimators are unbiased, consistent, and asymptotically normal.\cite{vanbreuk}

\subsubsection{Relationship with two-sample U-statistics} \label{ss: ratio}
When all clusters feature the same number of individuals so that $n_{ic} = n$ for all $i$ and $c$, and endpoints are equally weighted, the mixed model estimator $\bar{Y}_{i\cdot \cdot \cdot}$ in Equation (\ref{eq: gls_mean}) reduces to the simple mean of the global win fractions in the $i^{th}$ arm,
\begin{equation}
\bar{Y}_{i \cdot \cdot \cdot} = \frac{1}{N_i} \sum_{c=1}^{C_i} \sum_{j=1}^{n} \bar{Y}_{icj\cdot}.
\label{eq: ratio}
\end{equation}
Expansion of Equation (\ref{eq: ratio}) according to the global win fraction definitions within Section \ref{ss: gwf} provides, 
\begin{equation}
\hat{\theta} = \frac{1}{K} \sum_{k=1}^{K} \left\lbrace \frac{1}{N_1 N_0} \sum_{c=1}^{C_1} \sum_{c'=1}^{C_0} \sum_{j=1}^{n_{1c}} \sum_{j'=1}^{n_{0c'}} H(X_{1cjk} - X_{0c'j'k}) \right\rbrace
\label{eq: plugin}
\end{equation}
suggesting that $\hat{\theta}$ is equivalent to the simple mean of $K$ clustered, two-sample U-statistics when individuals are equally weighted. 

Obuchowski\cite{obuchowski} developed the same estimator (\ref{eq: plugin}) for a single endpoint ($K=1$) by extending DeLong et al's AUC estimators,\cite{delong} or equivalently Sen's structural component estimators,\cite{sen1960} to the clustered setting. Brunner et al\cite{multibrunmun} established the unbiasedness, consistency, and multivariate normality of multiple win probabilities with independent observations, and Rubarth et al\cite{konietschke} established a similar result for clustered factorial designs, even when the $w_{ic}$ defined above are incorporated. From these results, it follows for cluster randomized trials that the global win probability, equal to the mean of the multiple win probabilities, is asymptotically $\hat{\theta} \sim \mathcal{N}(\theta, \text{Var}(\hat{\theta}))$.

\subsubsection{Equivalence to mean difference in rank-sums} \label{ss: mdrs}
Again, we consider the scenario where equal weight is assigned to each individual and endpoint to demonstrate the relationship between the difference in mean rank-sums originally considered by O'Brien\cite{obrien} and the global win probability estimator within cluster trials. That is, we let $\hat{\theta} = \sum_{c} \sum_{j} \bar{Y}_{1cj\cdot} / N_1$ where $\bar{Y}_{1cj\cdot}$ is the global win fraction of the $cj^{th}$ treatment individual.

From the rank-based form of the win fractions presented previously in Equation (\ref{eq: rank_y}) and the fact that the sum of the group-specific ranks $\sum_{c} \sum_{j} G_{1cjk} = N_1 (N_1 + 1)/2$, with some algebra it follows that,
$$ \hat{\theta} = \frac{1}{N_0} \left[ \frac{\bar{R}_{1\cdot\cdot\cdot}}{K} - \frac{(N_1 + 1)}{2} \right] $$
where $R_{1cj\cdot} = \sum_{k} R_{1cjk}$ is the rank-sum for the $cj^{th}$ treatment individual and $\bar{R}_{1\cdot\cdot\cdot} = \sum_{c} \sum_{j} R_{1cj\cdot} / N_1$ is the mean treatment rank-sum. Similarly for the control arm, 
$$ 1-\hat{\theta} = \frac{1}{N_1} \left[ \frac{\bar{R}_{0\cdot\cdot\cdot}}{K} - \frac{(N_0 + 1)}{2} \right].$$
Thus, the mean difference in rank-sums is a linear transformation of the global win probability estimator with 
\begin{equation*}
\bar{R}_{1\cdot\cdot\cdot} - \bar{R}_{0\cdot\cdot\cdot} = NK(\hat{\theta} - 0.5)
\end{equation*}
and $\widehat{\text{Var}}(\bar{R}_{1\cdot\cdot\cdot} - \bar{R}_{0\cdot\cdot\cdot}) = N^2 K^2 \widehat{\text{Var}}(\hat{\theta})$.

\subsection{Interval estimators and hypothesis tests} \label{ss: i_est}

By relying on asymptotic normality, a large-sample $(1-\alpha) \times 100\%$ confidence interval for the global win probability is provided by
$$ \hat{\theta} \mp z_{\alpha/2} \sqrt{\widehat{\text{Var}}(\hat{\theta})}$$
where $z_{\alpha/2}$ is the upper $\alpha/2$ quantile of $\mathcal{N}(0,1)$. For smaller samples, $z_{\alpha/2}$ may be substituted with the corresponding Student's $t$ critical value with df degrees of freedom, $t_{\alpha/2, \text{df}}$. Commonly, $\text{df}=C-2$ where $C$ is the total number of clusters, but there is no unique way of specifying the degrees of freedom of a mixed model.\cite{crts} The corresponding test of the null hypothesis of no global treatment effect, $H_0: \theta = 0.5$, can be assessed using the test statistic,
$$ T = \frac{\hat{\theta} - 0.5}{\sqrt{\widehat{\text{Var}}(\hat{\theta})}}$$
which is distributed according to $\mathcal{N}(0,1)$ for large samples or approximately $t_{\text{df}}$ for small samples. Using results from Section \ref{ss: mdrs}, the nonparametric rank-sum test is analogous to a test of the global win probability with $H_0: \theta = 0.5$ as,
$$ T = \frac{\hat{\theta} - 0.5}{\sqrt{\widehat{\text{Var}}(\hat{\theta})}} =  \frac{\bar{R}_{1\cdot\cdot\cdot} - \bar{R}_{0\cdot\cdot\cdot}}{\sqrt{\widehat{\text{Var}}(\bar{R}_{1\cdot\cdot\cdot} - \bar{R}_{0\cdot\cdot\cdot})}}.$$

A logit transformation may also be applied to improve behaviour for a small number of clusters or extreme values of $\hat{\theta}$. The lower and upper bounds of the large-sample $(1-\alpha) \times 100\%$ logit interval, $[L_{\text{logit}}, U_{\text{logit}}]$, are obtained respectively as,
$$ L_{\text{logit}} = \frac{\exp(\ell)}{1 + \exp(\ell)} \>,\> U_{\text{logit}} = \frac{\exp(u)}{1+\exp(u)} $$
where
$$ \ell, u = \ln \frac{\hat{\theta}}{1-\hat{\theta}} \mp z_{\alpha/2} \frac{\sqrt{\widehat{\text{Var}}(\hat{\theta})}}{\hat{\theta} (1-\hat{\theta})}.$$
The null hypothesis of no treatment effect, $H_0: \text{logit}(\theta) = 0$, can also be assessed using the test statistic,
$$ T_{\text{logit}} = \frac{\text{logit}(\hat{\theta})}{\hat{\theta}(1-\hat{\theta})/\sqrt{\widehat{\text{Var}}(\hat{\theta})}} $$
which is distributed according to $\mathcal{N}(0,1)$ for large samples.

\section{Simulation studies}
\subsection{Objectives and evaluation}
This simulation study assesses the performance of the proposed interval estimators for the global win probability and their corresponding hypothesis tests of $H_0: \theta = 0.5$ across a range of cluster trial designs. Performance metrics of interest are the coverage probability, balance of left and right tail error rates, type I error rates, and power. Emphasis is placed on the evaluation of interval estimators as their performance provides a combined summary of the quality of the proposed point and variance estimators. 

Empirical coverage probability (ECP) is estimated as the proportion of confidence intervals containing the true global win probability $\theta$, while left (right) tail error rates are estimated as the proportion of lower (upper) bounds greater than (less than) $\theta$. The tail error ratio (TER), defined as the ratio of the left tail error rate to the right, is reported with a TER of 1 suggesting balance. Empirical type I error rates and power are estimated as the proportion of confidence intervals excluding the null value of 0.5, or the empirical rejection rate (ERR), when $\theta$ is null and non-null, respectively.

Reported ECPs are considered acceptable if they fall within approximately two standard errors of the specified nominal rate. Specifically, $95\%$ confidence intervals are desired, and each scenario is replicated $B=5,000$ times so that the acceptable range of the empirical coverage probability is $0.95 \mp 1.95 \sqrt{0.95(0.05)/5000}$ or $94.4\%$ to $95.6\%$. The corresponding acceptable range for the empirical type I error rate is $(1-0.956) = 0.044$ to $(1-0.944) = 0.056$. 

\subsection{Scenarios and data generation}
Our simulation study employed a factorial design, evaluating 192 scenarios in total. Focus is restricted to the most common 1:1 allocation scheme with an equal number of clusters in each arm so that $C_0 = C_1 = C/2$, where the total number of clusters $C = 10, 20, 30$ and $50$. As discussed in Section \ref{ss: ratio}, the mixed model point estimator is equivalent to the unweighted mean of the global win fractions when cluster sizes are equal. Thus, two scenarios are considered: (i) a reference case with equal cluster sizes so that $n_{ic} = n = 30$ for all $i$ and $c$; and (ii) a realistic case with unequal, average cluster sizes of $\bar{n} = 30$ individuals per cluster. For unequal cluster sizes, $n=40$ individuals were generated per cluster and deleted completely at random according to a Bernoulli distribution with deletion probability 0.25 to obtain the desired $\bar{n}.$

Sample size parameters correspond to the median number of clusters randomized and median number of individuals per cluster reported by several reviews of cluster trials. Among cluster trials in primary care, Eldridge et al reported a median of 34 clusters randomized with a median cluster size of 32 individuals (IQR = 9 to 82).\cite{eldridge} Kahan et al reported a median of 25 clusters (IQR = 15 to 44) with a median cluster size of 34 individuals (IQR = 14 to 94).\cite{kahan} Ivers et al reported a median of 21 clusters (IQR = 12 to 52) with a median cluster size of 34 individuals (IQR = 13 to 89).\cite{ivers} 


Our simulation study considers only $K=2$ endpoints. Since our methods are particularly advantageous for ordinal endpoints lacking meaningful units or endpoints with different scales,\cite{zou_pharma} we let $X_{0cj1} \sim \text{Binomial}(4, 0.5)$ and $X_{0cj2} \sim \text{Binomial}(6, 0.5)$ to reflect commonly encountered 5- and 7-point Likert scales. Parameters within the treatment arm were then chosen so that $\theta = 0.5, 0.56, 0.64, 0.71$, or to yield null, small, medium, and large effect sizes based on relationships with Cohen's effect size when endpoints are normal.\cite{effectsizes} Endpoint effects $\theta_k$ could be homogeneous such that $\theta_1 = \theta_2 = \theta$ or heterogeneous such that $\theta_1 \neq \theta_2$. Heterogeneous effects were set to $\theta_1 = (\theta - 0.03)$ and $\theta_2 = (\theta + 0.03)$ so that a ``small" difference existed between the two effects. The true global win probability is $\theta = (\theta_1 + \theta_2)/2.$

Let $\Omega$ represent the $2\times2$ within-subject correlation matrix,
\begin{equation*}
\Omega = \left[ 
\begin{array}{cc}
1 & \omega_{12} \\
\omega_{12} & 1 
\end{array}
 \right]
\end{equation*}
with off-diagonal entries $\omega_{12}=\omega_{21}$ equal to the pairwise correlation of the two endpoints on their original scale, \textit{i.e.}, $\text{Corr}(X_{icj1}, X_{icj2})$. The pairwise correlation was varied from weak to strong, with $\omega_{12} = 0.3, 0.5,$ and $0.8$.

Let $\Phi$ represent the $2\times2$ matrix of intracluster correlations,
\begin{equation*}
\Phi = \left[ 
\begin{array}{cc}
\phi_{11} & \phi_{12} \\
\phi_{12} & \phi_{22}
\end{array}
 \right]
\end{equation*}
with diagonal entries $\phi_{kk}$ equal to the ICC of the $k^{th}$ endpoint, \textit{i.e.}, $\text{Corr}(X_{icjk}, X_{icj'k})$ where $j \neq j'$, and off-diagonal entries $\phi_{12}=\phi_{21}$ equal to the cross-subject cross-endpoint within-cluster correlation, \textit{i.e.}, $\text{Corr}(X_{icjk}, X_{icj'k'})$ where $j \neq j'$ and $k' \neq k$. Values of the intracluster correlation are generally small. A review of estimated intracluster correlations within primary care trials reported that 99\% of ICCs were less than 0.10.\cite{icc} Focusing on heterogeneous endpoint ICCs, the ICCs were set to values of $\phi_{11} = 0.1$, $\phi_{22} = 0.05$, and $\phi_{12}=0.025$. 

The within-cluster correlation matrix is then provided by $\mathbf{R}_{ic} = \mathbf{I} \otimes (\Omega - \Phi) + \mathbf{J} \otimes \Phi$ where $\mathbf{I}$ is the $n_{ic} \times n_{ic}$ identity matrix, $\mathbf{J}$ is the $n_{ic} \times n_{ic}$ matrix of ones, and $\otimes$ denotes the Kronecker product.\cite{wang} In other words, $\mathbf{R}_{ic}$ is a block diagonal matrix with $\Omega$ repeated on the diagonal $n_{ic}$ times and each off-diagonal entry equal to $\Phi$. For a cluster with three individuals, \textit{i.e.}, $n_{ic} = 3$, 
\begin{equation*}
\mathbf{R}_{ic} = \left[ \begin{array}{cc|cc|cc}
1 & \omega_{12} & \phi_{11} & \phi_{12} & \phi_{11} & \phi_{12} \\
\omega_{12} & 1 & \phi_{12} & \phi_{22} & \phi_{12} & \phi_{22} \\ \hline
\phi_{11} & \phi_{12} & 1 & \omega_{12} & \phi_{11} & \phi_{12} \\
\phi_{12} & \phi_{22} & \omega_{12} & 1 & \phi_{12} & \phi_{22} \\ \hline
\phi_{11} & \phi_{12} & \phi_{11} & \phi_{12} & 1 & \omega_{12}  \\
\phi_{12} & \phi_{22} & \phi_{12} & \phi_{22} & \omega_{12} & 1 
\end{array} \right].
\end{equation*}
The components of $\mathbf{R}_{ic}$ were assumed to be the same for both arms, and clusters are independent within- and between-arms. 

Individual-level bivariate ordinal responses were generated independently by arm and cluster using the ``mean mapping algorithm" as implemented in SAS \texttt{PROC IML} by Wicklin.\cite{wicklin} In essence, realizations from a bivariate normal distribution with mean zero and intermediate correlation matrix $\tilde{\mathbf{R}}_{ic}$ are generated.  Quantiles of the univariate standard normal distribution are then used to convert the normal variates into ordinal variates with the desired marginal distributions $F_{ik}$ and correlation matrix $\mathbf{R}_{ic}$. Root-finding algorithms are required to identify the intermediate correlation or entries of $\tilde{\mathbf{R}}_{ic}$ which yield the desired correlation upon discretization.

\subsection{Results}
All performance metrics are reported in Tables \ref{tab: id1} and \ref{tab: logit1} for $C=10, 20$ total clusters and the identity and logit transformed intervals, respectively, and Tables \ref{tab: id2} and \ref{tab: logit2} for $C=30, 50$ total clusters. Results were similar regardless of whether effects were homogeneous or heterogeneous and clusters were equally or unequally sized.

For equal cluster sizes, empirical coverage probability (ECP) fell within the acceptable range of 94.4\% to 95.6\% for all scenarios and the desired type I error rate of $\alpha = 0.05$ was well-maintained. For unequal cluster sizes, ECP sometimes exceeded the upper acceptable bound of 95.6\%. This occurred primarily when there were fewer clusters per arm ($C=10$ or $20$) and effects were small to moderate. In the most extreme case, the ECP was $96\%$, suggesting that the interval estimators are at most slightly conservative. The nominal type I error rate $\alpha=0.05$ also appeared to be reasonably maintained for unequal cluster sizes.

For both unequal and equal cluster sizes, the tail error ratio (TER) was almost always greater than 1, suggesting that the left tail error rate has a tendency to exceed the right tail error rate. Discrepancy between tail errors appeared to increase as the global win probability $\theta$ increased or the number of clusters $C$ decreased. Use of the logit transformation improved balance of the tail errors considerably compared to the untransformed confidence interval while producing similar ECPs and ERRs. 

Increasing the pairwise correlation of the endpoints appeared to decrease the power to detect a global treatment effect, as should be expected. However, methods appeared generally powerful. Approximately 75\% power to detect a ``moderate" global effect $(\theta = 0.64)$ was achieved with as few as $C_i=5$ clusters per arm, while 80\% power to detect a ``small" global effect $(\theta = 0.56)$ required approximately $C_i=25$ clusters per arm.

\begin{table}[]
\centering 
\caption{Simulation results for the \textit{identity} interval estimator with $C=10, 20$ total clusters and equal allocation for increasing values of the global win probability $\theta$ and endpoint correlation $\omega_{12}$.}
\begin{tabular}{rrrrrrrrrrrrrrr}
\toprule
\multicolumn{1}{c}{} & \multicolumn{1}{c}{} & \multicolumn{1}{c}{} & \multicolumn{6}{c}{Homogeneous effects ($\theta_1 =\theta_2 = \theta$)}                         & \multicolumn{6}{c}{Heterogeneous effects ($\theta_{1} \neq \theta_{2}$)}                       \\ \cmidrule(lr){1-3}\cmidrule(lr){4-9}\cmidrule(lr){10-15} 
\multicolumn{3}{c}{Parameters}                                     & \multicolumn{3}{c}{Unequal $n_{ic}$} & \multicolumn{3}{c}{Equal $n$} & \multicolumn{3}{c}{Unequal $n_{ic}$} & \multicolumn{3}{c}{Equal $n$} \\ 
$C$                  & $\omega_{12}$        & $\theta$             & ECP     & TER     & ERR     & ECP    & TER    & ERR     & ECP     & TER     & ERR     & ECP    & TER    & ERR     \\ \cmidrule(lr){1-3}\cmidrule(lr){4-6}\cmidrule(lr){7-9}\cmidrule(lr){10-12}\cmidrule(lr){13-15}
10  & 0.3           & 0.50     & 95.1       & 0.93       & 4.9        & 94.8     & 1.18    & 5.2      & 95.1       & 0.88       & 4.9        & 94.9     & 1.02    & 5.1      \\
    &               & 0.56     & 94.8       & 1.04       & 22.4       & 94.3     & 1.44    & 21.1     & 95.0       & 1.16       & 22.1       & 94.7     & 1.34    & 21.0     \\
    &               & 0.64     & 95.1       & 1.53       & 78.8       & 94.7     & 1.47    & 78.8     & 95.3       & 1.30       & 78.5       & 94.6     & 1.58    & 78.7     \\
    &               & 0.71     & 94.6       & 1.79       & 98.7       & 94.5     & 2.03    & 99.0     & 94.8       & 1.93       & 98.9       & 94.5     & 2.04    & 99.0     \\
    & 0.5           & 0.50     & 95.3       & 0.93       & 4.7        & 94.7     & 1.09    & 5.3      & 95.0       & 0.97       & 5.0        & 94.8     & 1.10    & 5.2      \\
    &               & 0.56     & 94.8       & 1.25       & 20.5       & 94.8     & 1.40    & 20.6     & 95.0       & 1.05       & 21.5       & 94.9     & 1.30    & 20.5     \\
    &               & 0.64     & 95.1       & 1.63       & 77.7       & 94.6     & 1.56    & 76.7     & 94.8       & 1.42       & 77.1       & 94.5     & 1.63    & 76.7     \\
    &               & 0.71     & 95.1       & 1.87       & 98.5       & 94.6     & 1.90    & 98.5     & 95.0       & 1.96       & 98.6       & 94.6     & 1.92    & 98.7     \\
    & 0.8           & 0.50     & 95.5       & 1.08       & 4.5        & 95.0     & 1.19    & 5.0      & 95.5       & 1.05       & 4.5        & 94.8     & 1.02    & 5.2      \\
    &               & 0.56     & 95.2       & 1.26       & 19.6       & 94.9     & 1.37    & 19.5     & 95.1       & 1.30       & 20.3       & 94.9     & 1.28    & 19.4     \\
    &               & 0.64     & 94.9       & 1.53       & 75.2       & 94.7     & 1.61    & 74.8     & 95.3       & 1.35       & 74.9       & 95.0     & 1.50    & 74.4     \\
    &               & 0.71     & 94.9       & 2.21       & 98.1       & 95.1     & 2.25    & 98.0     & 94.7       & 2.22       & 98.0       & 95.2     & 2.18    & 98.0     \\ \cmidrule(lr){1-3}\cmidrule(lr){4-6}\cmidrule(lr){7-9}\cmidrule(lr){10-12}\cmidrule(lr){13-15}
20  & 0.3           & 0.50     & 95.2       & 1.10       & 4.8        & 94.6     & 1.20    & 5.4      & 94.8       & 1.03       & 5.2        & 94.7     & 1.11    & 5.3      \\
    &               & 0.56     & 95.0       & 1.19       & 42.4       & 94.7     & 1.35    & 41.5     & 95.4       & 1.09       & 42.4       & 94.8     & 1.43    & 41.0     \\
    &               & 0.64     & 95.1       & 1.52       & 98.5       & 94.7     & 1.68    & 98.8     & 95.1       & 1.46       & 98.5       & 94.9     & 1.67    & 98.8     \\
    &               & 0.71     & 95.1       & 2.00       & 100.0      & 94.8     & 2.08    & 100.0    & 94.8       & 1.82       & 100.0      & 94.6     & 1.98    & 100.0    \\
    & 0.5           & 0.50     & 95.3       & 0.86       & 4.7        & 94.6     & 1.19    & 5.4      & 95.2       & 1.06       & 4.8        & 94.6     & 1.23    & 5.4      \\
    &               & 0.56     & 95.6       & 1.08       & 41.5       & 94.5     & 1.38    & 39.9     & 95.4       & 1.05       & 41.3       & 94.8     & 1.44    & 39.9     \\
    &               & 0.64     & 95.1       & 1.74       & 98.2       & 94.7     & 1.77    & 98.3     & 94.8       & 1.76       & 98.2       & 94.9     & 1.53    & 98.3     \\
    &               & 0.71     & 94.9       & 2.56       & 100.0      & 94.6     & 2.19    & 100.0    & 95.2       & 1.78       & 100.0      & 94.4     & 1.99    & 100.0    \\
    & 0.8           & 0.50     & 95.2       & 0.94       & 4.8        & 94.5     & 1.22    & 5.5      & 95.2       & 0.83       & 4.8        & 94.6     & 1.23    & 5.4      \\
    &               & 0.56     & 95.5       & 1.07       & 39.3       & 94.6     & 1.40    & 38.0     & 95.0       & 1.10       & 38.8       & 94.5     & 1.28    & 37.8     \\
    &               & 0.64     & 95.3       & 1.61       & 97.7       & 94.5     & 1.62    & 97.8     & 95.2       & 1.42       & 97.7       & 94.7     & 1.67    & 97.7     \\
    &               & 0.71     & 94.9       & 1.68       & 100.0      & 94.6     & 2.18    & 100.0    & 95.1       & 1.44       & 100.0      & 94.4     & 1.86    & 100.0    \\ \bottomrule
\end{tabular}
\flushleft
\footnotesize{ECP: Empirical Coverage Probability, TER: Tail Error Ratio, ERR: Empirical Rejection Rate.}
\label{tab: id1}
\end{table}

\begin{table}[]
\centering 
\caption{Simulation results for the \textit{identity} interval estimator with $C=30, 50$ total clusters and equal allocation for increasing values of the global win probability $\theta$ and endpoint correlation $\omega_{12}$.}
\begin{tabular}{rrrrrrrrrrrrrrr}
\toprule
\multicolumn{1}{c}{} & \multicolumn{1}{c}{} & \multicolumn{1}{c}{} & \multicolumn{6}{c}{Homogeneous effects ($\theta_1 = \theta_2 = \theta$)}                         & \multicolumn{6}{c}{Heterogeneous effects ($\theta_1 \neq \theta_2$)}                       \\ \cmidrule(lr){1-3}\cmidrule(lr){4-9}\cmidrule(lr){10-15} 
\multicolumn{3}{c}{Parameters}                                     & \multicolumn{3}{c}{Unequal $n_{ic}$} & \multicolumn{3}{c}{Equal $n$} & \multicolumn{3}{c}{Unequal $n_{ic}$} & \multicolumn{3}{c}{Equal $n$} \\ 
$C$                  & $\omega_{12}$        & $\theta$             & ECP     & TER     & ERR     & ECP    & TER    & ERR     & ECP     & TER     & ERR     & ECP    & TER    & ERR     \\ \cmidrule(lr){1-3}\cmidrule(lr){4-6}\cmidrule(lr){7-9}\cmidrule(lr){10-12}\cmidrule(lr){13-15}
30  & 0.3           & 0.50     & 94.6       & 0.92       & 5.4        & 94.9     & 1.03    & 5.1      & 95.0       & 0.97       & 5.0        & 95.1     & 1.11    & 4.9      \\
    &               & 0.56     & 95.1       & 1.15       & 58.3       & 95.1     & 1.22    & 58.3     & 95.3       & 1.11       & 58.1       & 95.2     & 1.15    & 58.6     \\
    &               & 0.64     & 95.2       & 1.48       & 99.9       & 94.9     & 1.40    & 99.9     & 95.2       & 1.28       & 99.9       & 95.1     & 1.48    & 99.9     \\
    &               & 0.71     & 94.8       & 1.70       & 100.0      & 95.0     & 1.75    & 100.0    & 94.8       & 1.55       & 100.0      & 94.9     & 1.67    & 100.0    \\
    & 0.5           & 0.50     & 94.7       & 0.98       & 5.3        & 94.9     & 1.06    & 5.1      & 94.7       & 0.97       & 5.3        & 95.0     & 0.98    & 5.0      \\
    &               & 0.56     & 95.2       & 1.24       & 56.7       & 95.1     & 1.15    & 56.3     & 95.3       & 1.19       & 56.6       & 95.1     & 1.18    & 56.6     \\
    &               & 0.64     & 95.1       & 1.65       & 99.9       & 94.9     & 1.52    & 99.9     & 94.7       & 1.32       & 99.9       & 95.1     & 1.50    & 99.9     \\
    &               & 0.71     & 94.5       & 1.68       & 100.0      & 94.9     & 1.64    & 100.0    & 94.8       & 1.57       & 100.0      & 94.9     & 1.65    & 100.0    \\
    & 0.8           & 0.50     & 95.0       & 1.07       & 5.0        & 94.9     & 0.96    & 5.1      & 95.1       & 0.98       & 4.9        & 94.9     & 1.00    & 5.2      \\
    &               & 0.56     & 95.0       & 1.16       & 54.4       & 95.4     & 1.11    & 54.4     & 95.2       & 1.14       & 54.4       & 95.3     & 1.05    & 54.2     \\
    &               & 0.64     & 95.0       & 1.55       & 99.8       & 94.9     & 1.25    & 99.8     & 95.1       & 1.32       & 99.9       & 95.1     & 1.31    & 99.8     \\
    &               & 0.71     & 94.7       & 1.75       & 100.0      & 95.1     & 1.66    & 100.0    & 94.6       & 1.63       & 100.0      & 95.1     & 1.56    & 100.0    \\ \cmidrule(lr){1-3}\cmidrule(lr){4-6}\cmidrule(lr){7-9}\cmidrule(lr){10-12}\cmidrule(lr){13-15}
50  & 0.3           & 0.50     & 95.4       & 1.16       & 4.6        & 95.5     & 1.06    & 4.5      & 95.0       & 1.19       & 5.0        & 95.4     & 0.98    & 4.6      \\
    &               & 0.56     & 95.2       & 1.44       & 81.7       & 95.5     & 1.18    & 81.3     & 94.9       & 1.24       & 81.6       & 95.3     & 1.18    & 81.7     \\
    &               & 0.64     & 94.6       & 1.52       & 100.0      & 95.3     & 1.36    & 100.0    & 94.9       & 1.43       & 100.0      & 95.3     & 1.41    & 100.0    \\
    &               & 0.71     & 95.0       & 1.60       & 100.0      & 95.4     & 1.66    & 100.0    & 94.9       & 1.79       & 100.0      & 95.2     & 1.57    & 100.0    \\
    & 0.5           & 0.50     & 95.3       & 1.10       & 4.7        & 95.0     & 1.16    & 5.0      & 94.9       & 1.06       & 5.1        & 95.0     & 1.12    & 5.0      \\
    &               & 0.56     & 94.8       & 1.33       & 80.5       & 95.0     & 1.36    & 79.9     & 95.1       & 1.19       & 80.2       & 94.8     & 1.17    & 80.3     \\
    &               & 0.64     & 94.7       & 1.63       & 100.0      & 94.8     & 1.51    & 100.0    & 95.3       & 1.48       & 100.0      & 94.9     & 1.51    & 100.0    \\
    &               & 0.71     & 94.7       & 1.78       & 100.0      & 94.9     & 1.67    & 100.0    & 94.9       & 1.84       & 100.0      & 94.6     & 1.63    & 100.0    \\
    & 0.8           & 0.50     & 95.4       & 1.30       & 4.6        & 94.3     & 1.17    & 5.7      & 95.1       & 1.09       & 4.9        & 94.5     & 1.17    & 5.5      \\
    &               & 0.56     & 95.0       & 1.18       & 77.0       & 94.4     & 1.24    & 77.0     & 95.2       & 1.29       & 77.5       & 94.7     & 1.18    & 77.0     \\
    &               & 0.64     & 94.5       & 1.43       & 100.0      & 94.7     & 1.47    & 100.0    & 95.0       & 1.26       & 100.0      & 94.5     & 1.45    & 100.0    \\
    &               & 0.71     & 94.8       & 1.55       & 100.0      & 94.6     & 1.51    & 100.0    & 95.0       & 1.46       & 100.0      & 94.8     & 1.54    & 100.0      \\ \bottomrule
\end{tabular}
\flushleft
\footnotesize{ECP: Empirical Coverage Probability, TER: Tail Error Ratio, ERR: Empirical Rejection Rate.}
\label{tab: id2}
\end{table}

\begin{table}[]
\centering 
\caption{Simulation results for the \textit{logit} interval estimator with $C=10, 20$ total clusters and equal allocation for increasing values of the global win probability $\theta$ and endpoint correlation $\omega_{12}$.}
\begin{tabular}{rrrrrrrrrrrrrrr}
\toprule
\multicolumn{1}{c}{} & \multicolumn{1}{c}{} & \multicolumn{1}{c}{} & \multicolumn{6}{c}{Homogeneous effects ($\theta_1 = \theta_2 = \theta$)}                              & \multicolumn{6}{c}{Heterogeneous effects ($\theta_1 \neq \theta_2$)}                            \\ \cmidrule(lr){1-3}\cmidrule(lr){4-9}\cmidrule(lr){10-15}
\multicolumn{3}{c}{Parameters}                                     & \multicolumn{3}{c}{Unequal $n_{ic}$} & \multicolumn{3}{c}{Equal $n$} & \multicolumn{3}{c}{Unequal $n_{ic}$} & \multicolumn{3}{c}{Equal $n$} \\  
$C$                  & $\omega_{12}$        & $\theta$             & ECP        & TER        & ERR        & ECP      & TER     & ERR      & ECP        & TER        & ERR        & ECP      & TER     & ERR      \\ \cmidrule(lr){1-3}\cmidrule(lr){4-6}\cmidrule(lr){7-9}\cmidrule(lr){10-12}\cmidrule(lr){13-15}
10         & 0.3           & 0.50     & 95.5                & 0.94 & 4.5   & 95.2      & 1.15 & 4.8   & 95.5                  & 0.88 & 4.5   & 95.2      & 1.07 & 4.8   \\
           &               & 0.56     & 95.3                & 0.84 & 20.8  & 94.9      & 1.21 & 19.9  & 95.4                  & 1.02 & 20.5  & 95.0      & 1.20 & 19.4  \\
           &               & 0.64     & 95.6                & 1.00 & 76.4  & 94.9      & 1.16 & 76.4  & 95.8                  & 0.84 & 76.1  & 95.1      & 1.26 & 76.1  \\
           &               & 0.71     & 95.2                & 1.00 & 98.4  & 95.1      & 1.24 & 98.6  & 95.3                  & 1.10 & 98.4  & 95.1      & 1.26 & 98.5  \\
           & 0.5           & 0.50     & 95.7                & 0.94 & 4.3   & 95.1      & 1.08 & 4.9   & 95.5                  & 0.94 & 4.4   & 95.2      & 1.06 & 4.8   \\
           &               & 0.56     & 95.2                & 1.10 & 19.0  & 95.1      & 1.24 & 19.3  & 95.3                  & 0.93 & 20.0  & 95.1      & 1.16 & 19.0  \\
           &               & 0.64     & 95.7                & 1.07 & 75.1  & 95.0      & 1.18 & 74.0  & 95.4                  & 0.97 & 74.5  & 95.0      & 1.10 & 74.3  \\
           &               & 0.71     & 95.6                & 1.11 & 98.1  & 95.1      & 1.19 & 98.0  & 95.4                  & 1.02 & 98.1  & 95.2      & 1.06 & 98.1  \\
           & 0.8           & 0.50     & 96.0                & 1.07 & 4.0   & 95.4      & 1.17 & 4.6   & 95.8                  & 1.06 & 4.2   & 95.2      & 1.02 & 4.8   \\
           &               & 0.56     & 95.6                & 1.06 & 18.6  & 95.4      & 1.16 & 18.2  & 95.6                  & 1.12 & 19.0  & 95.2      & 1.14 & 18.0  \\
           &               & 0.64     & 95.3                & 1.05 & 72.4  & 95.3      & 1.10 & 72.1  & 95.7                  & 0.97 & 72.3  & 95.2      & 1.20 & 71.5  \\
           &               & 0.71     & 95.5                & 1.27 & 97.3  & 95.4      & 1.21 & 97.3  & 95.2                  & 1.28 & 97.1  & 95.5      & 1.26 & 97.3  \\ \cmidrule(lr){1-3}\cmidrule(lr){4-6}\cmidrule(lr){7-9}\cmidrule(lr){10-12}\cmidrule(lr){13-15}
20         & 0.3           & 0.50     & 95.6                & 1.12 & 4.4   & 94.8      & 1.24 & 5.2   & 95.1                  & 1.01 & 4.9   & 94.9      & 1.12 & 5.1   \\
           &               & 0.56     & 95.1                & 1.11 & 41.5  & 94.9      & 1.21 & 40.5  & 95.4                  & 1.03 & 41.5  & 94.9      & 1.34 & 40.0  \\
           &               & 0.64     & 95.3                & 1.17 & 98.4  & 95.1      & 1.29 & 98.6  & 95.3                  & 1.17 & 98.3  & 95.3      & 1.28 & 98.6  \\
           &               & 0.71     & 95.4                & 1.48 & 100.0 & 94.9      & 1.35 & 100.0 & 95.2                  & 1.09 & 100.0 & 95.1      & 1.25 & 100.0 \\
           & 0.5           & 0.50     & 95.5                & 0.88 & 4.5   & 94.7      & 1.21 & 5.3   & 95.4                  & 1.05 & 4.6   & 94.9      & 1.23 & 5.1   \\
           &               & 0.56     & 95.8                & 0.98 & 40.5  & 94.8      & 1.27 & 38.7  & 95.6                  & 0.94 & 40.3  & 94.9      & 1.38 & 38.8  \\
           &               & 0.64     & 95.4                & 1.27 & 98.0  & 95.0      & 1.34 & 98.2  & 95.2                  & 1.22 & 98.1  & 95.1      & 1.20 & 98.2  \\
           &               & 0.71     & 95.4                & 1.52 & 100.0 & 94.9      & 1.42 & 100.0 & 95.3                  & 1.18 & 100.0 & 94.9      & 1.35 & 100.0 \\
           & 0.8           & 0.50     & 95.3                & 0.93 & 4.7   & 94.7      & 1.20 & 5.3   & 95.4                  & 0.81 & 4.6   & 94.7      & 1.21 & 5.3   \\
           &               & 0.56     & 95.7                & 0.95 & 38.3  & 94.7      & 1.28 & 36.9  & 95.2                  & 0.97 & 37.8  & 94.8      & 1.17 & 36.8  \\
           &               & 0.64     & 95.4                & 1.22 & 97.5  & 94.7      & 1.25 & 97.6  & 95.3                  & 1.12 & 97.5  & 95.0      & 1.27 & 97.6  \\
           &               & 0.71     & 95.4                & 1.04 & 100.0 & 94.5      & 1.29 & 100.0 & 95.2                  & 0.90 & 100.0 & 94.5      & 1.12 & 100.0 \\ \bottomrule
\end{tabular}
\flushleft
\footnotesize{ECP: Empirical Coverage Probability, TER: Tail Error Ratio, ERR: Empirical Rejection Rate.}
\label{tab: logit1}
\end{table}

\begin{table}[]
\centering 
\caption{Simulation results for the \textit{logit} interval estimator with $C=30, 50$ total clusters and equal allocation for increasing values of the global win probability $\theta$ and endpoint correlation $\omega_{12}$.}
\begin{tabular}{rrrrrrrrrrrrrrr}
\toprule
\multicolumn{1}{c}{} & \multicolumn{1}{c}{} & \multicolumn{1}{c}{} & \multicolumn{6}{c}{Homogeneous effects ($\theta_1 = \theta_2 = \theta$)}                              & \multicolumn{6}{c}{Heterogeneous effects $(\theta_1 \neq \theta_2)$}                            \\ \cmidrule(lr){1-3}\cmidrule(lr){4-9}\cmidrule(lr){10-15}
\multicolumn{3}{c}{Parameters}                                     & \multicolumn{3}{c}{Unequal $n_{ic}$} & \multicolumn{3}{c}{Equal $n$} & \multicolumn{3}{c}{Unequal $n_{ic}$} & \multicolumn{3}{c}{Equal $n$} \\  
$C$                  & $\omega_{12}$        & $\theta$             & ECP        & TER        & ERR        & ECP      & TER     & ERR      & ECP        & TER        & ERR        & ECP      & TER     & ERR      \\ \cmidrule(lr){1-3}\cmidrule(lr){4-6}\cmidrule(lr){7-9}\cmidrule(lr){10-12}\cmidrule(lr){13-15}
30         & 0.3           & 0.50     & 94.7                & 0.93 & 5.3   & 95.1      & 1.06 & 4.9   & 95.2                  & 0.99 & 4.8   & 95.2      & 1.09 & 4.8   \\
           &               & 0.56     & 95.3                & 1.01 & 57.6  & 95.2      & 1.12 & 57.6  & 95.5                  & 1.03 & 57.3  & 95.3      & 1.07 & 58.0  \\
           &               & 0.64     & 95.4                & 1.16 & 99.9  & 95.2      & 1.06 & 99.9  & 95.3                  & 1.12 & 99.9  & 95.0      & 1.11 & 99.9  \\
           &               & 0.71     & 95.0                & 1.17 & 100.0 & 95.2      & 1.14 & 100.0 & 95.1                  & 0.99 & 100.0 & 95.1      & 1.17 & 100.0 \\
           & 0.5           & 0.50     & 95.0                & 0.95 & 5.0   & 95.2      & 1.00 & 4.8   & 94.8                  & 1.02 & 5.1   & 95.1      & 0.98 & 4.9   \\
           &               & 0.56     & 95.5                & 1.08 & 56.0  & 95.0      & 1.07 & 55.7  & 95.4                  & 1.07 & 55.9  & 95.3      & 1.04 & 55.9  \\
           &               & 0.64     & 95.3                & 1.22 & 99.9  & 95.1      & 1.15 & 99.9  & 94.7                  & 1.09 & 99.9  & 95.1      & 1.15 & 99.9  \\
           &               & 0.71     & 94.7                & 1.14 & 100.0 & 95.2      & 1.13 & 100.0 & 95.0                  & 1.06 & 100.0 & 95.2      & 1.05 & 100.0 \\
           & 0.8           & 0.50     & 95.1                & 1.04 & 4.9   & 95.1      & 1.00 & 4.9   & 95.3                  & 0.96 & 4.7   & 95.1      & 0.98 & 4.9   \\
           &               & 0.56     & 95.1                & 1.02 & 53.6  & 95.5      & 0.99 & 53.5  & 95.4                  & 1.03 & 53.8  & 95.5      & 0.94 & 53.6  \\
           &               & 0.64     & 95.3                & 1.17 & 99.8  & 95.1      & 1.01 & 99.8  & 95.3                  & 0.96 & 99.9  & 95.2      & 1.03 & 99.8  \\
           &               & 0.71     & 94.9                & 1.16 & 100.0 & 95.1      & 1.08 & 100.0 & 94.7                  & 1.10 & 100.0 & 95.1      & 1.08 & 100.0 \\ \cmidrule(lr){1-3}\cmidrule(lr){4-6}\cmidrule(lr){7-9}\cmidrule(lr){10-12}\cmidrule(lr){13-15}
50         & 0.3           & 0.50     & 95.4                & 1.18 & 4.6   & 95.5      & 1.08 & 4.5   & 95.1                  & 1.20 & 4.9   & 95.5      & 1.02 & 4.5   \\
           &               & 0.56     & 95.3                & 1.28 & 81.5  & 95.5      & 1.12 & 81.2  & 95.0                  & 1.15 & 81.4  & 95.5      & 1.08 & 81.3  \\
           &               & 0.64     & 94.8                & 1.21 & 100.0 & 95.5      & 1.14 & 100.0 & 95.1                  & 1.21 & 100.0 & 95.4      & 1.12 & 100.0 \\
           &               & 0.71     & 95.1                & 1.21 & 100.0 & 95.5      & 1.13 & 100.0 & 95.1                  & 1.31 & 100.0 & 95.4      & 1.21 & 100.0 \\
           & 0.5           & 0.50     & 95.4                & 1.11 & 4.6   & 95.2      & 1.16 & 4.8   & 94.9                  & 1.06 & 5.1   & 95.0      & 1.12 & 4.9   \\
           &               & 0.56     & 94.9                & 1.25 & 80.2  & 95.1      & 1.23 & 79.5  & 95.3                  & 1.11 & 79.7  & 94.9      & 1.11 & 80.1  \\
           &               & 0.64     & 94.8                & 1.36 & 100.0 & 95.1      & 1.29 & 100.0 & 95.5                  & 1.16 & 100.0 & 95.0      & 1.24 & 100.0 \\
           &               & 0.71     & 94.8                & 1.24 & 100.0 & 95.1      & 1.26 & 100.0 & 95.2                  & 1.23 & 100.0 & 95.0      & 1.31 & 100.0 \\
           & 0.8           & 0.50     & 95.5                & 1.30 & 4.5   & 94.4      & 1.16 & 5.6   & 95.2                  & 1.05 & 4.9   & 94.6      & 1.13 & 5.4   \\
           &               & 0.56     & 95.2                & 1.07 & 76.7  & 94.6      & 1.13 & 76.6  & 95.3                  & 1.17 & 77.1  & 94.8      & 1.11 & 76.8  \\
           &               & 0.64     & 94.7                & 1.15 & 100.0 & 94.8      & 1.22 & 100.0 & 95.0                  & 1.06 & 100.0 & 94.7      & 1.18 & 100.0 \\
           &               & 0.71     & 95.0                & 1.23 & 100.0 & 94.9      & 1.11 & 100.0 & 95.2                  & 1.01 & 100.0 & 95.1      & 1.10 & 100.0 \\ \bottomrule
\end{tabular}
\flushleft
\footnotesize{ECP: Empirical Coverage Probability, TER: Tail Error Ratio, ERR: Empirical Rejection Rate.}
\label{tab: logit2}
\end{table}

\section{Case study: SHARE}
\subsection{Motivation}
The Sexual Health and Relationships: Safe, Happy and Responsible (SHARE) trial aimed to determine whether an experimental sexual education curriculum reduced unsafe sex among students when compared to the existing curriculum.\cite{share} Schools in Scotland were randomized as clusters to be trained according to the experimental SHARE curriculum or not. This case study uses a subset of the trial data originally accessed from the Harvard Dataverse, with a link provided in the Appendix alongside reproducible R and SAS code.

Two endpoints are considered here: (1) knowledge, an ordinal endpoint ranging from -8 to 8, or least sexual health knowledge to most; and (2) activity, a binary endpoint taking the value $1$ if the student was sexually active during follow-up and $0$ otherwise. Suppose that because the intervention is curriculum-based, investigators prioritize improvement in knowledge over a decrease in activity, though both effects are desired. Investigators wish to estimate the global treatment effect, assigning weights $w_1 = 0.7$ to knowledge and $w_2 = 0.3$ to activity, respectively. Traditional composite approaches may dichotomize knowledge to combine it with activity, potentially resulting in ceiling or floor effects, or ignore differences in priority. 

This case study demonstrates how weights can be incorporated within the global win probability to provide an interpretable estimate of the global treatment effect that respects differences in both endpoint priority and type. Transformation to alternative win measures is also demonstrated.

\subsection{Descriptives by endpoint}
We consider $C_0 = 12$ and $C_1 = 13$ clusters assigned to conventional and experimental curriculum, respectively. Only students with a completely observed bivariate response were retained for analysis, resulting in an average of 193 students per cluster or $N=4,823$ total students ($N_0 = 2,474$,  $N_1=2,349$). Table \ref{tab: share_desc} provides a summary of the observed endpoints, their corresponding win fractions, and correlation. 

\begin{table}[]
\centering
\caption{Summary of observed knowledge and activity scores and win fractions for $N_0 = 2,474$ and  $N_1=2,349$ individuals receiving conventional and experimental curriculum within SHARE.}
\label{tab: share_desc}
\begin{tabular}{llrrcc}
\toprule
              &                                    & \multicolumn{2}{c}{\textbf{Mean (SD)}}                                       & \multicolumn{2}{c}{\textbf{Correlation}} \\ \cmidrule(lr){3-4}\cmidrule(lr){5-6}
              & \textbf{Endpoint} & \multicolumn{1}{c}{Conventional} & \multicolumn{1}{c}{Experimental} & Knowledge           & Activity          \\ \midrule
Obs. scores   & Knowledge                          & 4.10 (2.36)                      & 4.75 (2.28)                      & 1.00                & 0.08              \\
              & Activity                           & 0.27 (0.44)                      & 0.27 (0.44)                      & 0.08                & 1.00              \\ \midrule
Win fractions & Knowledge                          & 0.42 (0.28)                      & 0.58 (0.28)                      & 1.00                & -0.08              \\
              & Activity                           & 0.50 (0.22)                      & 0.50 (0.22)                       & -0.08                & 1.00   
              \\ \bottomrule
\end{tabular}
\end{table}

The mean treatment win fraction was approximately 57.9\% for knowledge and 50.1\% for activity. These estimates suggest no difference in activity but a small benefit of the SHARE curriculum compared to the conventional curriculum with respect to sexual health knowledge. The win fraction ICCs of knowledge and activity were both 0.028, and the corresponding observed ICCs were 0.031 and 0.028.

\subsection{Estimates and interpretation}
The use of SAS \texttt{PROC MIXED} to estimate the two-level mixed model presented in Equation (\ref{mod}) provides $\hat{\beta}_1 = 0.104$ with $\widehat{\text{SE}}(\hat{\beta}_1) = 0.017$ on $\text{df}=23$ degrees of freedom, $\hat{\sigma}^2_{\alpha} \approx 0.002$ and $\hat{\sigma}^2_{\varepsilon} \approx 0.040$. The corresponding weighted global win probability estimate is $\hat{\theta} = 1.104/2 = 0.552$ with $\widehat{\text{SE}}(\hat{\theta}) = 0.017$ and a logit transformed 95\% confidence interval of $(0.517 \text{ to } 0.587)$. The estimated ICC of the global win fractions was $0.037$.

Formally, the estimated probability that a student receiving the experimental curriculum responds better than a student receiving the existing curriculum is 55.2\% (95\% CI: 0.517 to 0.587), on average, with respect to the two weighted endpoints of sexual knowledge and activity. Since the null value $\theta_0 = 0.5$ is excluded from the 95\% confidence interval, there is statistically significant evidence at the $\alpha=0.05$ level to reject the null hypothesis of no weighted global treatment effect. For reference, if equal weights are assigned to each endpoint, the estimated global win probability is 0.537 (95\% CI: 0.506 to 0.568).

Alternatively, the estimated global win difference is $\hat{\Delta} = 2(0.552) - 1 = 0.104$ ($\widehat{\text{SE}}(\hat{\Delta}) = 2(0.017) = 0.034$; 95\% CI: 0.037 to 0.171), suggesting that the probability of winning on treatment is approximately 10.4 basis points greater than that on control. The estimated win odds are $\hat{\lambda} = 0.552/(1-0.552) = 1.23$ ($\widehat{\text{SE}}(\hat{\lambda}) = 0.017/(0.552 \times 0.448)=0.069$; 95\% CI: 1.095 to 1.365), suggesting that the treatment win probability is approximately 23\% greater than the control win probability. Noting that the null values of the win difference and win odds are $\Delta_0 = 0$ and $\lambda_0 = 1$, respectively, the 95\% confidence intervals suggest evidence against the null hypothesis of no global treatment effect at the $\alpha = 0.05$ level.

\section{Discussion}
The design and analysis of trials with multiple endpoints face statistical complexities,\cite{fda_endpts} which are further exacerbated by intracluster correlation induced by cluster randomization. To address these challenges, we present interval estimation and hypothesis testing methods for a nonparametric global treatment effect, referred to as the global win probability, for cluster randomized trials with multiple endpoints. The global win probability directly quantifies the objective of most trials which is to determine if patients receiving treatment have better overall health outcomes than those on control or the current standard of care. The global win probability only compares the ordering of responses, rather than their size or difference, and is applicable to any endpoint that can be ranked including binary, ordinal, count, and continuous endpoints. Estimation is also robust to monotonic transformation, unlike mean-based methods which may yield differing or even conflicting results pre- and post-transformation.

To estimate the global win probability, a single rank-based global win fraction is constructed for each individual within the cluster trial as the within-subject mean of their endpoint-specific win fractions. Global win fractions are interpretable at the individual-level, unlike alternative composite measures, as the average proportion of responses within the comparator arm exceeded by a given individual. Weights may also be incorporated within construction of the global win fractions to reflect differences in endpoint utility or priority. The mixed model estimation framework previously introduced by Zou\cite{zou} for a single win probability is then easily applied to the univariate global win fractions to obtain point, variance, and interval estimators of the global win probability adjusted for intracluster correlation. The developed methods are simple as they bypass the need to consider complex correlation structures between-endpoint and within-cluster, and accessible as they may be implemented using standard statistical software as demonstrated by the R and SAS code provided in the Appendix.

Simulation results suggest that the developed methods are powerful and yield interval estimators that respect nominal coverage and type I error rates across a range of cluster randomized designs. Reported empirical coverage probabilities were close to the nominal confidence level of 95\% in most scenarios, though left tail error rates were predominantly larger than those of the right tail. Thus, its recommended that the logit-transformed interval estimators be used to achieve greater balance, particularly when few clusters are randomized or a large global treatment effect is expected. Performance did not appear to differ depending on whether endpoint-specific effects were homogeneous or heterogeneous, while power increased as the correlation between endpoints decreased. Thus, power to detect a global treatment effect may be increased by considering endpoints that capture non-overlapping aspects of disease to reduce correlation.

Since all endpoints are reframed in terms of ``wins," or better health, the multiple endpoints may be either positively or negatively correlated. This is not always true of alternative composite measures or global treatment effects. When effects are discordant, \textit{i.e.}, some endpoints exhibit benefit and others harm, the global win probability may be interpreted as a measure of the overall risk-benefit trade-off. When effects are concordant, \textit{i.e.}, all endpoints share similar direction and magnitude, power will be maximized as a result of the global treatment effect assumption. Either way, the win probability for each endpoint should always be investigated and reported descriptively to understand their influence on the global treatment effect. The inability of some composites to reflect differences in endpoint priority is also a commonly cited concern.\cite{rauch} Endpoint priority may differ when, \textit{e.g.}, endpoint utility, importance, or severity differs. The global win probability, and corresponding global win fractions, allow investigators to explicitly specify the desired contribution of each endpoint to the composite and global treatment effect. The global win probability is also applicable when priority ordering is ill-defined or endpoints are indeed equally important. 

If desired, the global win probability can also be directly transformed into other popular, alternative effect measures encountered in medicine. This includes win measures such as the win difference or win odds,\cite{buyse, winodds} or more traditional effects such as the risk difference or standardized mean difference.\cite{effectsizes} The global win probability is inspired by the nonparametric rank-sum test introduced by O'Brien for individually randomized trials with multiple disparate endpoints.\cite{obrien} As demonstrated here, the mean difference in rank-sums employed by this test is also a linear transformation of global win probability estimators.

The flexible mixed model framework employed also permits the consideration of more complex models. For example, Zou et al recently detailed how win fraction regression methods could be used to analyze individually randomized pre-post designs in a fashion analogous to ANCOVA.\cite{zou_pharma} A similar approach could be taken here by, \textit{e.g.}, regressing global win fractions at follow-up on global win fractions at baseline, potentially leading to additional boosts in power.\cite{edward} Stratified or minimized designs could also perhaps be accommodated through the inclusion of design covariates. It would also be worthwhile to investigate multivariate linear mixed models for the win fractions. These models would allow, \textit{e.g.}, the construction of a $K$-df test, use of more complex covariance structures, or assessment of global treatment effects over time. 

Another advantage of using the mixed model framework is that the form of sample size estimators is relatively simple. That is, the corresponding sample size for an individually randomized trial can be scaled by a design effect equal to a function of cluster size and the ICC, \textit{i.e.}, $\text{DE}=1 + (n-1) \rho$. Sample size formulas for a single win probability within individually randomized trials were recently provided by Zou et al.\cite{zou_contemp} However, parameters required for estimation are currently difficult to ascertain due to the novelty of win probability methods, \textit{i.e.}, estimates are not currently reported in the literature. Fortunately, relationships between endpoint distributions or alternative measures and the win probabilitiy are well-known, with a nice summary provided by Rahlfs and Zimmerman\cite{effectsizes}, for example. Thus, parametric treatment effect estimates reported in the literature may be transformed into endpoint-specific win probabilities and averaged to obtain an approximate global win probability for sample size estimation. Standard errors may also be obtained, see Shu \& Zou for example.\cite{shu} However, further investigation into the true relationship between the observed and win fraction ICCs is needed. Zou et al\cite{zou_pharma} suggested use of the observed pre-post correlation for sample size. Using the largest, conservative, endpoint ICC estimate may be a reasonable strategy, for now.

Several areas of future work would assist with making these win fraction methods viable in practice. First, accommodations for missing or censored responses are needed. The Heaviside function in Section \ref{ss: gwf}, for example, could be modified to incorporate censoring indicators. However, the resulting effect estimate may be dependent on the censoring distribution, requiring further adjustment. Adjustment or imputation techniques for missing responses should also be investigated. Second, the work of Zou et al\cite{zdk} on group sequential methods for cluster randomized trials with binary outcomes, for example, may serve as a basis for developing such methods for clustered win probability methods. Finally, the provided SAS and R code aims to assist with future implementation of the methods. However, the development of formal SAS macros and an R library would be ideal.

\section{Bibliography} \linespread{1}
\bibliography{citations}


\begin{appendix}

\clearpage
\section{SHARE} \label{app: SHARE}
Data from the SHARE trial can be downloaded from the Harvard Dataverse in tab-delimited format (\texttt{share.tab}) at \url{https://dataverse.harvard.edu/dataverse/crt}.
\subsection{R code}
\linespread{1}
\begin{verbatim}
library(dplyr) # Data manipulation
library(nlme)  # Mixed models

# IMPORT: SHARE data
share_raw <- read.table('share.tab', sep='\t', header=T)

# EXTRACT: identifiers and outcomes (Note: school=cluster)
share <- share_raw %>%
  select(arm, school, idno, 'knowledge'=kscore, 'active'=debut) %>%
  na.omit()

# EXTRACT: Number of students in each arm
N0 <- nrow(share[share$arm == 0,]) # Conventional
N1 <- nrow(share[share$arm == 1,]) # Experimental

# CONSTRUCT: overall and group ranks for each endpoint
# Note: increased knowledge = good, increased active = bad
ranks <- share %>%
  mutate(R1 = rank(knowledge), R2 = rank(-active)) %>%
  group_by(arm) %>%
  mutate(G1 = rank(knowledge), G2 = rank(-active))

# Endpoint weights
w1 = 0.7; w2 = 0.3

# CONSTRUCT: endpoint and global win fractions
winf <- ranks %>%
  mutate(Y1 = ifelse(arm == 0, (R1-G1)/N1, (R1-G1)/N0),
         Y2 = ifelse(arm == 0, (R2-G2)/N1, (R2-G2)/N0),
         YG = w1 * Y1+ w2* Y2)

# FIT: linear mixed model for global win fractions
modG <- lme(YG ~ arm, random = ~1 | school, data=winf)

# EXTRACT: Fixed effect estimates, their variance, and df
modG_fest  <- fixef(modG)
modG_fvar  <- vcov(modG)
modG_fdf   <- modG$fixDF$X  # Note: Arm df = C-2 by default

# EXTRACT: Random effect variance components
modG_rvar  <- matrix(as.numeric(VarCorr(modG)), ncol=2)

# CONSTRUCT: Global win probability point, variance, and ICC est
estG <- (modG_fest[2] + 1)/2
seG  <- sqrt(modG_fvar[2,2])
iccG <- modG_rvar[1,1] / (modG_rvar[1,1] + modG_rvar[2,1])

# CONSTRUCT: Global win probability interval estimates
alpha <- 0.05; t <- qt(1-alpha/2, modG_fdf[2])
untransform_ci <- estG + c(-1, 1) * t * seG
logit_lu <- log(estG/(1-estG)) + c(-1, 1) * t * seG/(estG*(1-estG))
logit_ci <- exp(logit_lu) / (1 + exp(logit_lu))

# REPORT: Estimates
report_point <- paste0('Est. global win probability = ', round(estG, 4),
                       ' (Est. SE = ', round(seG, 4), ', df = ', modG_fdf[2], ')')
report_icc <- paste0('Est. global ICC = ', round(iccG, 4))
report_uci <- paste0('95% untransformed confidence interval = (', 
                     round(untransform_ci, 4)[1], ', ', round(untransform_ci, 4)[2], ')')
report_lci <- paste0('95% logit confidence interval = (', 
                     round(logit_ci, 4)[1], ', ', round(logit_ci, 4)[2], ')')

cat(paste(report_point, report_uci, report_lci, report_icc, sep='\n'))
\end{verbatim}
\pagebreak

\subsection{SAS code}
\linespread{1}
\begin{verbatim}
PROC IMPORT DATAFILE="share.tab" OUT=share DBMS=DLM REPLACE; DELIMITER='09'x; RUN;

DATA share; SET share;
	knowledge = kscore;
	* Reverse code active since higher => worse;
	active = -1 * debut;
	* Exclude individuals with incomplete responses;
	IF (knowledge ne .) and (active ne .) THEN OUTPUT;
	KEEP arm school idno knowledge active;
RUN;

* Sort data by Arm (0 = Conventional, 1 = Experimental);
PROC SORT DATA=share; BY arm; RUN;

* Obtain number of individuals in each arm;
PROC FREQ DATA=share NOPRINT; 
TABLE arm / OUT=SampleSize(DROP=percent); RUN;

* Reverse Arm labels for win fraction denominator;
DATA Denominator; SET SampleSize; arm = 1 - arm; RUN;

* Sort for merge;
PROC SORT DATA=Denominator; BY arm; RUN;

* Overall (mid)ranks for each endpoint;
PROC RANK DATA=share OUT=O_ranks TIES=mean; 
	VAR knowledge active; RANKS O1 O2;
RUN;

* Group (mid)ranks for each endpoint;
PROC RANK DATA=share OUT=G_ranks TIES=mean; BY arm;
	VAR knowledge active; RANKS G1 G2;
RUN;

* Calculate endpoint and global win fractions;
DATA WinF; MERGE O_ranks G_ranks Denominator; BY arm;
	Y1 = (O1-G1)/COUNT;
	Y2 = (O2-G2)/COUNT; 
	YG = 0.7*Y1 + 0.3*Y2;
RUN;

* Fit linear mixed model for global win fractions;
PROC MIXED DATA=WinF NOITPRINT NOCLPRINT;
	CLASS arm school idno / REF=first;
	MODEL YG = arm / solution; RANDOM intercept / SUBJECT=school(arm);
	ODS OUTPUT SolutionF = FixEff(KEEP=Arm Estimate StdErr DF);
	ODS OUTPUT CovParms = CovParms;
RUN;

* Extract estimates;
DATA VarInt VarRes; SET CovParms;
	IF CovParm='Intercept' THEN OUTPUT VarInt;
	IF CovParm='Residual'  THEN OUTPUT VarRes;
	DROP CovParm Subject;
RUN;

DATA FixEff; SET FixEff; Beta1 = Estimate; IF Arm = 1 THEN OUTPUT; RUN;

DATA GlobalEstimates;
	MERGE FixEff 
	VarInt(RENAME=(Estimate=VarAlpha))
	VarRes(RENAME=(Estimate=VarEps));
	EstG = (Beta1 + 1)/2; SeG = StdErr; DfG = DF;
	t = tinv(1-0.05/2, DF);
	* Untransformed 95% confidence interval;
	L1 = EstG - t * SeG; U1 = EstG + t * SeG;
	* Logit 95% confidence interval;
	L2 = log(EstG/(1-EstG)) - t * SeG / (EstG * (1-EstG));
	U2 = log(EstG/(1-EstG)) + t * SeG / (EstG * (1-EstG));
	L2 = exp(L2) / (1 + exp(L2)); U2 = exp(U2) / (1 + exp(U2));
	* Intracluster correlation of global win fractions;
	IccG = VarAlpha / (VarAlpha + VarEps); 
	KEEP EstG SeG DfG L1 U1 L2 U2 IccG;
RUN;

PROC PRINT DATA=GlobalEstimates; Run;
\end{verbatim}
\end{appendix}

\end{document}